\documentstyle[12pt]{article}
\newcommand{\be}{\begin{equation}}
\newcommand{\ee}{\end{equation}}
\newcommand{\ba}{\begin{eqnarray}}
\newcommand{\ea}{\end{eqnarray}}
\newcommand{\baa}{\begin{eqnarray*}}
\newcommand{\eaa}{\end{eqnarray*}}
\newcommand{\bb}{\begin{thebibliography}}
\newcommand{\eb}{\end{thebibliography}}

\newcommand{\bi}[1]{\bibitem{#1}}
\begin {document}

\rightline{ TPR-99-15}

\begin{center}
{\Large \bf Sum rules for the T-odd fragmentation functions}\\
\vspace*{1cm}
 A. Sch{\"a}fer$^1$, O.V. Teryaev$^{1,2}$ \\
\vspace*{0.3cm}
{$^1$\it Institut f{\"u}r Theoretische Physik\\
Universit{\"a}t Regensburg\\
D-93040 Regensburg, Germany }\\
\vspace*{0.3cm}
{$^2$\it Bogoliubov Laboratory of Theoretical Physics\\
Joint Institute for Nuclear Research, Dubna\\
141980 Russia}\\
\vspace*{0.3cm}
\end{center}
\begin{abstract}

The conservation of the intrinsic transverse momentum during
parton fragmentation imposes non-trivial constraints
on T-odd fragmentation functions. These
significantly enhance the differences between the favoured and unfavoured
fragmentation functions, which could be relevant to understand the
azimuthal asymmetries of charged pion production observed recently by
the HERMES collaboration.

\end{abstract}

\newpage

The T-odd effects in partonic fragmentation \cite{EMT,Col}
provide a unique tool for the study of parton polarization.
The key point is that they generate a self-analyzing pattern, which
connects
the parton polarizations in the hard process
to angular asymmetries in the produced hadrons.

The fragmentation functions, being the cornerstone of such processes,
should be very different from the standard ones. Their characteristic
property is
that they require the interference of two probability amplitudes
with a relative phase, which excludes the
usual probabilistic interpretation.
This makes the non-perturbative modelling of such effects rather
difficult. One possibility is to use an effective
quark propagator incorporating the imaginary part \cite{Col},
another are model calculations in which the phase shift is
produced by the Breit-Wigner propagator for a
wide hadronic resonances \cite{JT,hand}.

The available experimental data are also not numerous at the moment.
One should note the measurement of longitudinal handedness \cite{hand}
and the correlation of the Collins fragmentation function \cite{corr}
in $e^+e^-$-annihilation by the DELPHI collaboration.
At the same time, the recent measurements of azimuthal asymmetries
in DIS by SMC \cite{bravar} and HERMES \cite{av}
show the need for an extended theoretical treatment. In a complete
analysis the effects are
represented as convolutions of various distribution and fragmentation
functions \cite{MT} (and possibly, T-odd {\it fracture}
functions \cite{OT99}), for which  one needs some phenomenological
inputs.

In this situation model independent constraints for T-odd
fragmentation functions are most welcome. In the present article
we present a
method to derive such
constraints, which is
based on the conservation of the intrinsic transverse momentum of
hadrons in partons. The resulting zero sum rule for the fragmentation
function $H_1^{\perp (1)}$ allows to understand the flavour dependence of
the azimuthal asymmetry observed at HERMES.

To be more specific, let us start
with the definition of the fragmentation functions \cite{Col}

 \begin{eqnarray}
\label{def}
& \int \frac{dx^- d^2 x_T}{12(2\pi)^3} \exp \left(iP^+\frac{x^-}{z}+
iP_T \frac{x_T}{z}
\right)\;
Tr_{Dirac} \gamma^\mu \sum\limits_{P,X}
 <0|\psi (0)|P, X><P,X| \bar \psi (x)|0> &\nonumber \\
& = D  (z, k_T, \mu^2) P^\mu &\nonumber \\
& \int \frac{dx^- d^2 x_T}{12(2\pi)^3} \exp \left(iP^+\frac{x^-}{z}+
iP_T \frac{x_T}{z}
\right)\;
Tr_{Dirac} i \sigma^{\mu \nu}
\sum\limits_{P,X}
 <0|\psi (0)|P, X><P,X| \bar \psi (x)|0> &\nonumber \\
& = H_1^{\perp}  (z, k_T, \mu^2) \frac {P^\mu k_T^\nu}{M} &\nonumber \\
\end{eqnarray}
Here $k_T$ and $M$ are the intrinsic transverse momentum of the
parton and a mass
parameter of order of the jet mass.
We have also given, for comparison, the expression for the
standard spin-averaged
fragmentation function $D$. The hadron and parton momenta are related
by $P^+=zk^+, P_T=-zk_T$

Let us immediately come to the physical interpretation.
For $D$ it is most straigtforward. We adopt the normalization
conditions of \cite{Col} (differing from that of \cite{MT} by a factor
of $z$) so that $P(z,k_T)=D(z,k_T)/z$ is the probability
density to find a specific hadron with the specified momentum.
The transverse momentum integrated probability density
is therefore
\be
\label{aver}
P(z)=\int d^2 P_T \frac{D(z,-\frac{P_T}{z})}{z}=
z\int d^2 k_T D(z,k_T) \equiv zD(z),
\ee
so that longitudinal momentum conservation takes the form
\be
\label{long}
\sum_h\int_0^1 dz z^2 D(z)=1.
\ee

Let us pass on to T-odd fragmentation function. Although there is no direct
probabilistic interpretation (the relative phase shift
between the matrix elements
\be
<0|\sigma^{\mu \nu} \psi (0)|P, X>, <P,X| \bar \psi (x)|0>
\ee
is actually crucial), it may still be considered as a quark spin dependent
term in the differential cross-section, so that the corresponding
 probability
density is  proportional to
\be
\frac{H_1^{\perp} (z,k_T)}{z} k_T \cos\phi
\ee
Here $\phi$ is the azimuthal angle with respect to the plane of the given
component of transverse momentum.
One can now immediately write the {\it transverse} momentum
conservation as

\be
\label{aver1}
\sum_h\int dz \int d^2 P_T P_T k_T cos^2 \phi
\frac{H_1^{\perp} (z,-\frac{P_T}{z})}{z} \sim
\sum_h \int dz z^2
\int d^2 k_T k_T^2 H_1^{\perp} (z,k_T)=0.
\ee
This equality leads to the conservation of each of the two components
of transverse momentum, due to the arbitrary choice of the angle
$\phi$.
The integration over $\phi$ is factored out.
The integrated quantity is then
proportional to the transverse momentum averaged
function \cite{MT}
\be
\label{aver2}
H_1^{\perp (1)}(z)=\int d^2 k_T \frac{k_T^2}{2M^2}
H_1^{\perp} (z,k_T).
\ee
It then follows that
\be
\label{sr}
\sum_h \int dz z^2 H_1^{\perp (1)} (z)=0,
\ee

which is  our main result.

Several additional comments might help to better understand this formula.

First, note that the $k_T$ oddness required by  
the T-odd nature was absolutely crucial in obtaining the result.
While the conservation of the transverse
momentum
components applies to all $k_T$-dependent
functions, it gives only non-trivial constraints
for {\it odd}
powers of $k_T$. The similar
contribution to $D(z, k_T)$ is zero just because it is an even function
of $k_T$. Therefore,
similar sum rules hold for all $k_T$-odd fragmentation functions.
Moreover,
similar sum rules can also be derived for functions which
do not requiring intrinsic $k_T$ at all, like
for the twist-3 T-odd fragmentation function $c_V$ \cite{OT96},
which describes the
fragmentation of unpolarized quarks into polarized baryons.

Second, the chirality of the function is actually inessential
in the presented derivation, as the latter does not make
direct reference to the energy-momentum operator, which is chiral
even. For the actual function $H_1^{\perp (1)}$ one should think
about momentum conservation for processes initiated by the
tensor quark currents, rather than about matrix elements of the
momentum operators.

The immediate consequence of our sum rule is a larger difference between
favored (for $z \sim 1$) and unfavored fragmentation.
If the favored one is positive for large $z$, there must be
compensating negative contribution at lower $z$. For the unfavored
fragmentation function which probes generally smaller $z$ values the 
resulting cancellations should then be more pronounced.
Therefore we expect that for all T-odd fragmentation functions the
contributions from non-leading parton fragmentation will be severely
suppressed compared to the normal fragmentation function $D(z)$.
The resulting asymmetries should be much smaller for non-leading
hadrons than for leading ones. This behaviour actually shows up in a
comparison of the double semi-inclusive $\pi^+$ and $\pi^-$
asymmetries as measured by HERMES \cite{HERd} with the  $\pi^+$ and $\pi^-$
azimuthal single spin asymmetries, measured by the same collaboration
\cite{av,HERs}.
\vskip.3in
{ \Large \bf Acknowledgements:}\\
\vskip.2in
We acknowledge the useful
discussions with H. Avakian and A.V. Efremov.
A.S. acknowledges the support from DFG and BMBF.
O.V.T. was supported by Erlangen-Regensburg Graduiertenkollegs and by DFG in
the framework of Heisenberg-Landau Program of JINR-Germany Collaboration.


\bb{99}

\bi{EMT}
A.V. Efremov, L. Mankiewicz, N.A. Tornqvist. Phys.Lett.B284:394-400,1992

\bi{Col} J.C. Collins, Nucl.Phys.B396:161-182,1993

\bi{JT}
R.L. Jaffe, Xue-min Jin, Ji-an Tang, Phys.Rev.D57:5920-5922,1998

\bi{hand}
A.V. Efremov, L.G. Tkatchev, L.S. Vertogradov,
DELPHI-note 94-11 PHYS 355, 1994.
Proc. 27-th Int. Conf. on HEP (Glasgow, 1994),
IOP Pub. Ltd, London 1995, p.875.

\bi{corr}
A.V. Efremov, O.G. Smirnova, L.G. Tkatchev,
Nucl.Phys.Proc.Suppl.74:49-52,1999
hep-ph/9812522

\bi{bravar} A. Bravar, Contribution to DIS-99 Conference.

\bibitem{av} A. Airapetian et al., Observation of a Single-Spin
Azimuthal Asymmetry in Semi-Inclusive Pion Electro-Production, 
hep-ex/9910062.

\bi{MT}
P.J. Mulders, R.D. Tangerman,
Nucl.Phys.B461:197-237,1996, Erratum-ibid.B484:538-540,1997.

\bi{OT99} O. Teryaev,
Proceedings of the Workshop
`Polarized Protons at High Energies -
Accelerator Challenges and Physics Opportunities`,17-20 May 1999
DESY - Hamburg, Germany.

\bibitem{OT96} O.V. Teryaev, In SPIN-96
Proceedings,
p. 594.
Edited by C.W. de Jager, T.J. Ketel,
P.J. Mulders, J.E. Oberski, M. Oskam-Tamboezer. World Scientific, 1997.

\bi{HERd}  K. Ackerstoff et al.,  Phys.Lett.B464:123,1999.

\bi{HERs}  HERMES collaboration, to be published.

\eb

\end{document}